\documentclass[aip,rsi,amsmath,amssymb,reprint]{revtex4-1}
\usepackage{graphicx}% Include figure files
\usepackage{dcolumn}% Align table columns on decimal point
\usepackage{bm}% bold math
\usepackage[utf8]{inputenc}
\usepackage[T1]{fontenc}
\usepackage{mathptmx}
\usepackage{amsmath}

\usepackage[thinc]{esdiff}
\usepackage[dvipsnames]{xcolor}

\usepackage{natbib}
\usepackage[breaklinks=true]{hyperref}
\usepackage{breakcites}

\begin{document}

\preprint{AIP/123-QED}

\title{An Air Suspension to Demonstrate the Properties of Torsion Balances with Fibres of Zero Length}

% \\with Forced Linebreak
% Force line breaks with \\

\author{C. Gettings}
\email{cgettings@star.sr.bham.ac.uk}
% \altaffiliation[Also at ]{Physics Department, XYZ University.}%Lines break automatically or can be forced with \\

\author{C. C. Speake}

% \email{Second.Author@institution.edu.}

\affiliation{ Institute for Gravitational Wave Astronomy, Astrophysics and Space
Research Group, School of Physics and Astronomy, University of Birmingham,
Edgbaston, Birmingham, B15 2TT, United Kingdom%\\This line break forced with \textbackslash\textbackslash
}

% \author{C. Author}
% \homepage{http://www.Second.institution.edu/~Charlie.Author.}
% \affiliation{%
% Second institution and/or address%\\This line break forced% with \\
% }%

\date{\today}

% It is always \today, today,
%  but any date may be explicitly specified

\begin{abstract}
We report on the design and characterisation of an air-bearing suspension that has been constructed to highlight the properties of torsion balances with fibres of zero length. A float is levitated on this suspension and its rotational and translational motion in the horizontal plane of the laboratory is controlled using magnetic actuators. We demonstrate the in-situ electromagnetic tuning of the float's centre-of-buoyancy to an accuracy of $\pm$ 0.3 mm, which was limited by the noise in the air bearing. The rotational stiffness of the float, which is approximately zero by design, was also measured. We compare the observed behaviour of the float with the predictions of a detailed model of the statics of the float-actuator system. Finally, we briefly discuss the application of these ideas and results to the construction of sensitive devices for the measurement of weak forces with short ranges.
\end{abstract}
%\begin{quotation}
%The ``lead paragraph'' is encapsulated with the \LaTeX\ 
%\verb+quotation+ environment and is formatted as a single paragraph before the first section heading. 
%(The \verb+quotation+ environment reverts to its usual meaning after the first sectioning command.) 
%Note that numbered references are allowed in the lead paragraph.
%The lead paragraph will only be found in an article being prepared for the journal \textit{Chaos}.
%\end{quotation}

\maketitle

\section{\label{sec:Intro}Introduction}

%:\protect\\ The line break was forced \lowercase{via} \textbackslash\textbackslash}

Torsion balances have a long history in experimental science, stretching
back to the time of Cavendish~\cite{Cav}. In modern physics torsion
balances are used extensively in tests of the inverse square law of
gravity~\cite{ISL1,ISL2}, Casimir force measurements~\cite{Casimir}
and tests of the weak equivalence principle~\cite{WEP}. The advantage
of the torsion balance is that well-manufactured fibres have very
low rotational stiffness giving a high sensitivity, and well-designed
balances can be made such that, to some degree, tilt or horizontal
acceleration due to seismic noise will not couple to rotational motion
of the suspended mass (or bob)~\cite{Pros}. Despite these advantages
the traditional torsion balance design has some limitations, particularly
when it is employed to detect forces within sub-millimetre ranges.
Due to the vertical distance of the centre-of-mass from the point
of attachment, horizontal accelerations, due to micro-seismic motion
for example, can couple strongly to the simple pendulum mode. This
makes control of the torsion bob difficult. Also at some level there
will always be some coupling of tilt to rotational motion~\cite{Vibr}. Issues also face low-frequency torsion pendulum experiments where ground vibration and other sources of Newtonian noise become increasingly problematic~\cite{McEA2017,Shimoda_2019}. Tilt-rotational mode coupling is also a concern for seismic inertial sensors~\cite{Mow_Lowry_2019}. In addition to weak force measurements, the co-location of the centres-of-mass and buoyancy of a suspended mass is a crucial feature of horizontal accelerometers and tiltmeters~\cite{TiltSpeake,GW}. This is currently achieved only by the adjustment of small masses such as lockable screws.

The goal then is to create a device that shares the advantages of
the torsion balance, but is not limited by the drawbacks mentioned
above, where necessary adjustments to the centre-of-buoyancy can be achieved accurately and remotely irrespective of the device's environment. In a previous paper~\cite{Zero}, we showed how the stiffness of the actuators acting on a levitated object (referred to as a float) could be tuned in-situ in such a way that the centre-of-buoyancy of the levitation system could be altered to lie at the centre-of-mass of the float, and that the rotational stiffness could be tuned, ideally, to zero. This could all be achieved whilst simultaneously controlling
the translational degrees of freedom. The centre-of-mass
is the point where inertial forces act, whereas the centre-of-buoyancy
is the location of the resultant of the forces that are applied to
levitate the float and control its position. In the general case the
centre-of-mass will not lie at the centre-of-buoyancy
due to manufacturing imperfections and so horizontal accelerations
and tilts will couple to the rotational mode of the device. The classical
torsion balance has an in-built low sensitivity to tilt and horizontal
acceleration as the centre-of-buoyancy of the torsion bob can lie
to a good approximation on the rotational axis. In our previous paper~\cite{Zero},
we presented some initial results of measurements of the tuning of
the period of oscillation of a float suspended by perfect diamagnetism
(superconductivity). In this current paper we focus on the demonstration
of the precise tuning of the centre-of-buoyancy of a float.

We have constructed an air suspension, referred to here as the air
bearing~\cite{Jack}, that levitates the float. The float is then
controlled in the horizontal plane of the laboratory by magnetic actuators
which consist of coil-magnet pairs. By changing the currents in the
coils we can tune the centre-of-buoyancy of the float. We present
measurements which support this concept, suggesting that torsion balances
with fibres of zero length can indeed be tuned in-situ to be rotationally
decoupled from ground tilt and horizontal accelerations. The actuators
were designed such that the rotational stiffness of the float was
nominally zero. Without further tuning the magnitude of the rotational
stiffness was experimentally found to be lower than that of the torsion
balance used in a recent determination of Newton's constant of gravitation~\cite{BigG}.

\section{\label{sec:Theory}Theory}

In our previous publication we derived expressions for the rotational
stiffness and centre-of-buoyancy shift for superconducting and electrostatic
suspensions~\cite{Zero}. Here we give the corresponding expressions
for a system that uses electromagnetic actuators.

\begin{figure}
\includegraphics[width=1\columnwidth]{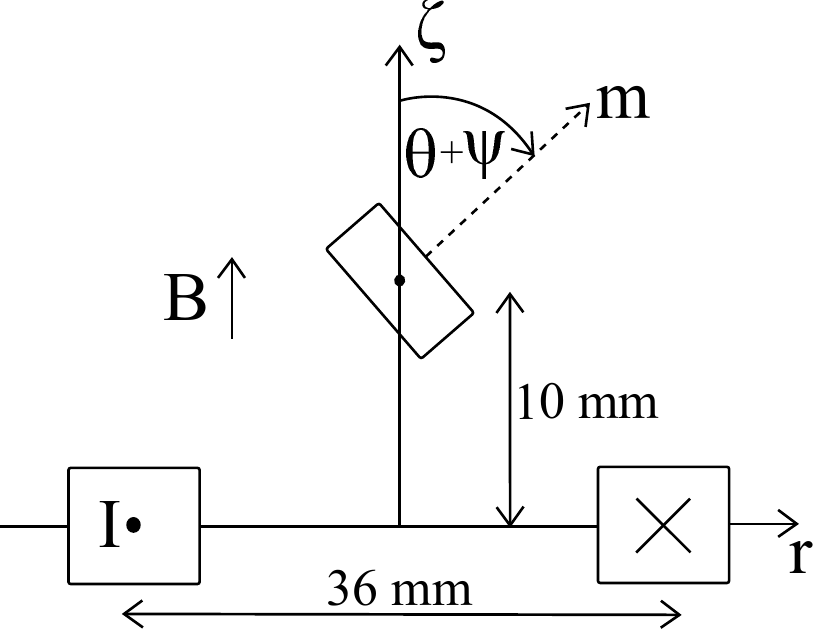}% Here is how to import EPS art
 \caption{\label{Basic} A schematic diagram of a magnet with
a magnetic dipole moment, $m$, at an angle, $\theta+\psi$, to a
magnetic field, $B$, produced by a coil with current $I$. The coil
has a cross-sectional area of $7.2\times10^{-5}$ m$^{2}$.}
\end{figure}

Consider Figure~\ref{Basic},
where a magnet with dipole moment, $m$, lies in a magnetic field
produced by a coil with current $I$. The magnetic
field on the axis of a coil of negligible cross-section
can be described by the following~\cite{Smythe}, 
\begin{equation}
B_{\zeta}=\frac{\mu_{0}R^{2}I}{2(R^{2}+\zeta^{2})^{\frac{3}{2}}},\label{Bz}
\end{equation}
where $\mu_{0}$ is the vacuum permeability constant, $R$ is the
radius of the coil and $\zeta$ is the axial distance between the
coil and magnet centres, as described by the coordinates in Figure~\ref{Basic}.
We can integrate this expression over the dimensions
of the real coils to find the field and its derivatives. If we assume
that the magnetic field is uniform over dimensions of the magnet,
we can write its potential energy in terms of its magnetic dipole
moment, 
\begin{equation}
U=-\vec{m}\cdot\vec{B}=-mB_{\zeta}\cos(\theta+\psi).\label{Energy}
\end{equation}
where $\theta$ is a fixed angle between the dipole moment and the
$\zeta$- axis and $\psi$ is a small angle whose mean is zero, as
indicated in Figure~\ref{Basic}. We ignore changes
in the axial force due to small radial displacements and rotations
of the magnet. For $\psi=0$ the force on the magnetic moment is
given by 
\begin{equation}
F=-\frac{\partial U}{\partial\zeta}=m\frac{\partial B_{\zeta}}{\partial\zeta}\cos\theta.\label{Force}
\end{equation}
Taking the magnetic field to be a maximum at the centre of the coil,
$\partial B_{\zeta}/\partial\zeta$ is negative, so for a magnetic
moment that is aligned with the field the force will be attractive
and reach a maximum negative value at some axial distance. The stiffness
in the $\zeta$ direction is given by 
\begin{equation}
k_{\zeta\zeta}=-\frac{\partial F}{\partial\zeta}=-m\frac{\partial^{2}B_{\zeta}}{\partial\zeta^{2}}\cos\theta.\label{LinStiff}
\end{equation}
In order for this stiffness to be positive, leading to a passively
stable system, we need the product of the cosine term and $\partial^{2}B_{\zeta}/\partial\zeta^{2}$
to be negative. At the peak force $\partial^{2}B_{\zeta}/\partial\zeta^{2}$
is zero, so in principle, we can choose the sign of the linear stiffness
by selecting the axial location of the magnet. If the magnet is closer
to the coil than the location of the peak force (where $\partial^{2}B_{\zeta}/\partial\zeta^{2}<0$)
and $\theta=0$, we can achieve a stable system. Equally we can achieve
a stable system by selecting a position of the magnet that is further
away from the coil than the peak force position and $\theta=\pi$.
Now consider the angular stiffness given as, again in the case where
$\psi=0$, 
\begin{equation}
k_{\theta\theta}=\frac{\partial^{2}U}{\partial\theta^{2}}=mB_{\zeta}\cos\theta.\label{eq:6-1}
\end{equation}
Clearly here the choice of $\theta$ will also determine the stability
of the system. We will see below, where we consider the stiffness
the whole float given by the actuators that control the float, that
it is advantageous to make the angular stiffness negative and we therefore
select $\theta=\pi$. If we desire a system that is stable for linear
motion, according to Equation~\ref{LinStiff}, we therefore need
to position the magnets further from the coils than the position of
maximum force. This is turn implies that the force between the magnet
and coil is repulsive. We should note that any unstable system can
be servo-controlled, however in practice servo control is more easily
achieved with an intrinsically stable system.

\begin{figure}
\includegraphics[width=1\columnwidth]{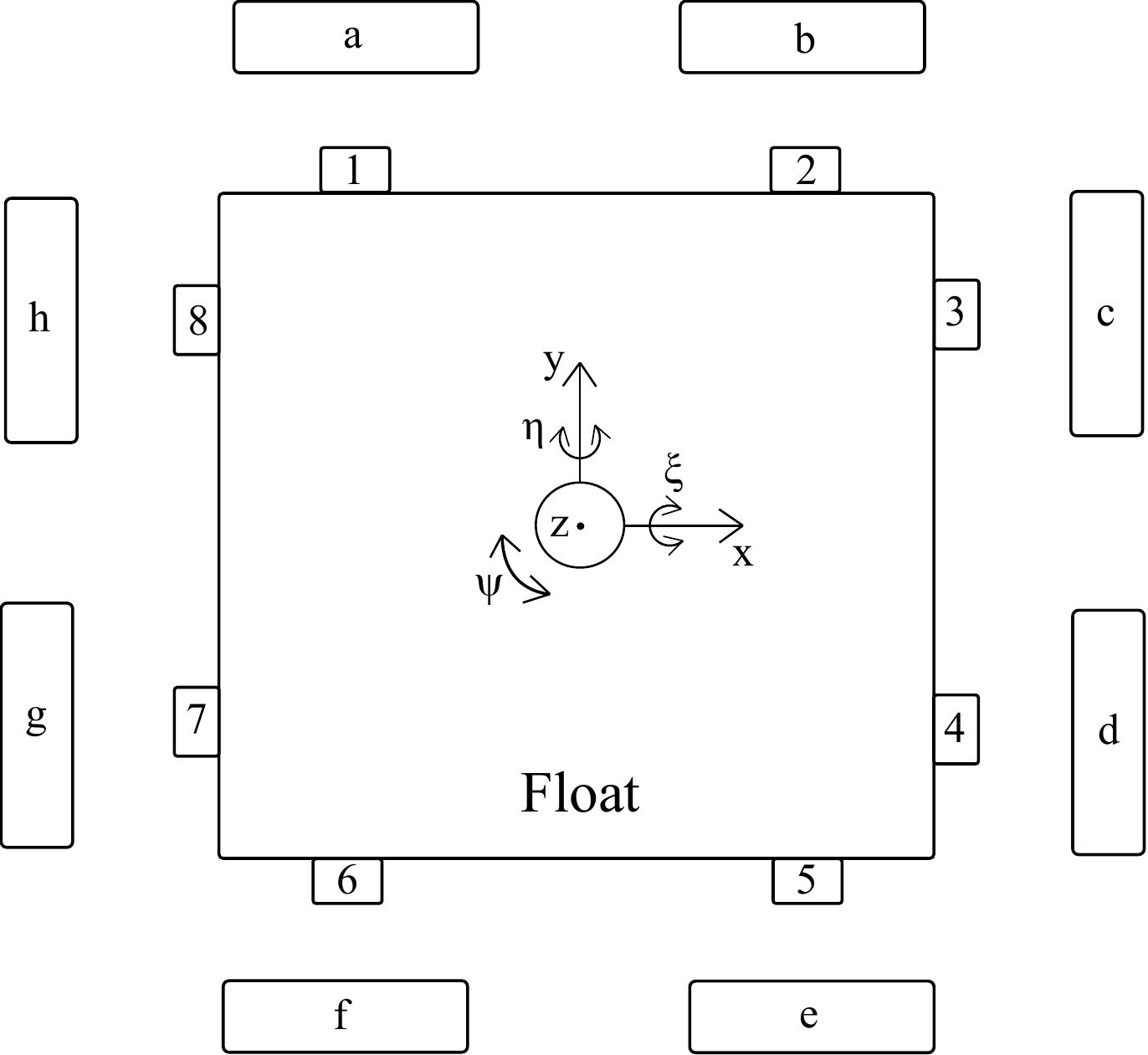}% Here is how to import EPS art
 \caption{\label{Setup} A plan view schematic drawing showing the float, coils
1-8 and magnets A-H. New global coordinates centred on the float are
defined: the x and y axes in the horizontal plane, and the z-axis
normal to this plane, with rotation around these axes.}
\end{figure}

\begin{figure}
	\includegraphics[width=1\columnwidth]{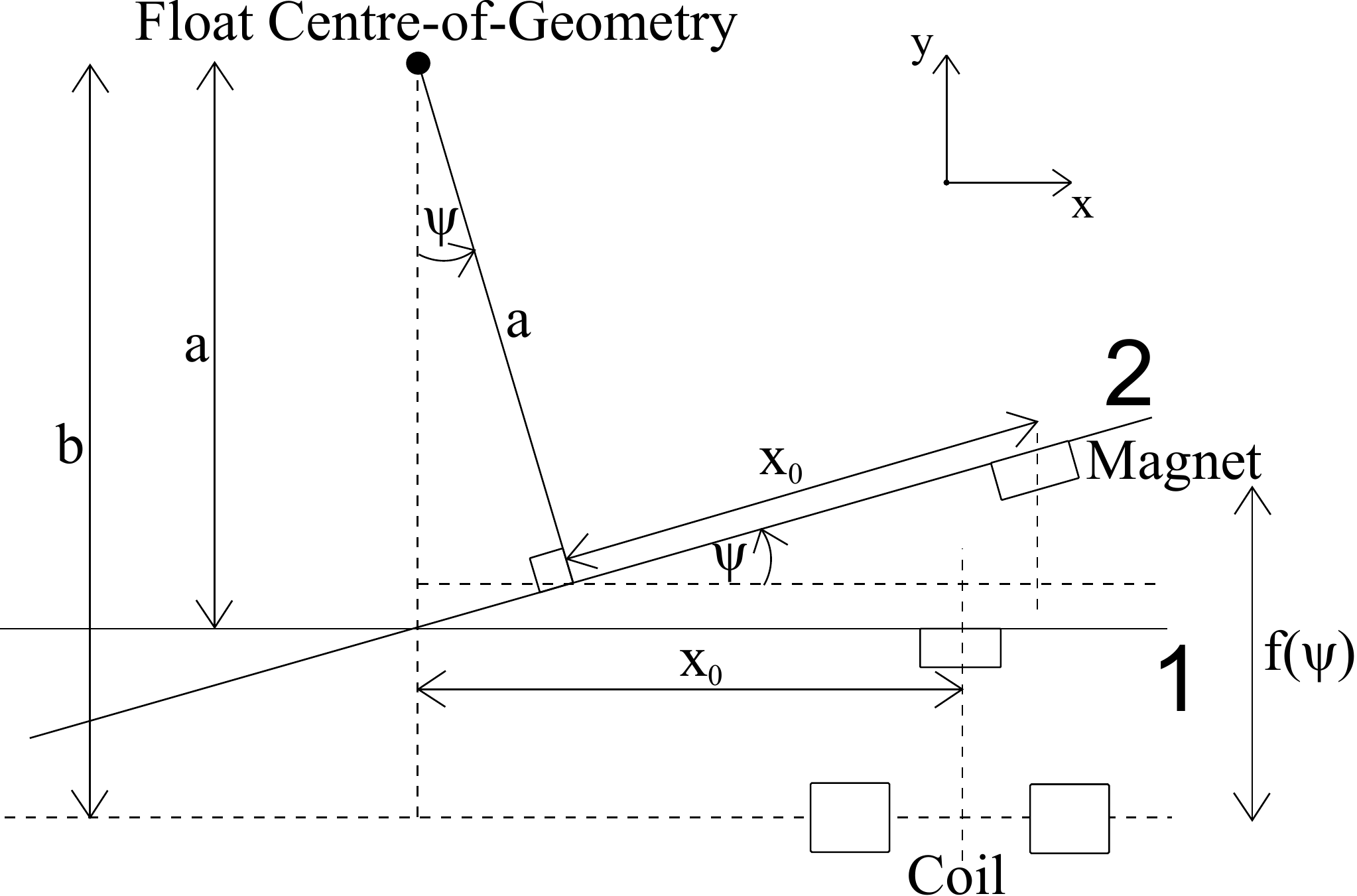}% Here is how to import EPS art
	\caption{\label{Separation} A diagram highlighting the geometry of the changing separation between magnet 5 and coil e (from Figure~\ref{Setup}) as described in Equation~\ref{eq:7}. As the float rotates around its geometric centre from position 1 to position 2, the separation between the magnet and coil, $f\left(\psi\right)$, is a function of the amount of rotation, $\psi$. This is independent of any separation change due to simple translation, $y$, along the y-axis as described in Equation~\ref{eq:5b}. The amount of rotation shown here is exaggerated for clarity, and as before we assume radial displacements along the x-axis are comparatively small and can be ignored. Here we have used a similar nomenclature as in Reference~13.}
\end{figure}

Now we consider our experimental setup with 8 such coil-magnet pairs
arranged around the float as shown in Figure~\ref{Setup}.
(with a new global coordinate system). We can now define the potential
energy of a single magnet/coil pair in terms of the coordinates and angles
given in Figure~\ref{Setup}, say for magnet 5 and coil e. We define the separation of the magnet from its opposing coil that is due to the rotation of the float, $\psi$, as $f\left(\psi\right)$. The separation due to its simple translation is defined as $y$, which is along the y-axis in Figure~\ref{Setup}. Figure~\ref{Separation} 
highlights the changing separation of this magnet and coil and the relevant geometric terms. We can therefore define 
\begin{equation}
\zeta=f\left(\psi\right)+y,\label{eq:5b}
\end{equation}
for this magnet/coil pair, and it can be easily shown from Figure~\ref{Separation} that 
\begin{equation}
f\left(\psi\right)=b-a\cos\psi+x_{0}\sin\psi.\label{eq:7}
\end{equation}
The potential energy of the magnet/coil pair can then be written,
\begin{equation}
U=-mB\left(f(\psi)+y\right)\cos\left(\theta+\psi\right),\label{eq:6}
\end{equation}
where we have dropped the subscript $\zeta$ on the magnetic field. This can be applied to any magnet/coil pair with a suitable change in coordinates. We have used a similar nomenclature as in Reference~13 and, on comparing Equation~\ref{eq:5b} with Equation~16 for $g(\psi)$ in Reference~13, we note that the magnetic/coil actuators are qualitatively different from the superconducting and electrostatic actuators previously discussed. This is because the points of application of the forces to the float (the magnets) by the coils now move with the rotation of the float rather than being defined by the fixed location of the coils. It is convenient to express the magnetic field as a Taylor series around the equilibrium position where we define the equilibrium spacing between the magnet and coil
as $g_{0}$. The magnetic field can then be written, 
\begin{multline}
B=B_{0}+\left(\frac{dB}{d\zeta}\right)_{g_{0}}\cdot(f\left(\psi\right)+y-g_{0})\\+\frac{1}{2}\left(\frac{d^{2}B}{d\zeta^{2}}\right)_{g_{0}}\cdot\left(f\left(\psi\right)+y-g_{0}\right)^{2}.\label{eq:8-1}
\end{multline}
By substituting this into Equation~\ref{eq:6} and setting $y=0$
we can find the contribution to the total rotational stiffness of
the float from one (the $i^{th}$) coil, 
\begin{equation}
K_{\psi\psi}^{i}=\frac{d^{2}U}{d\psi^{2}}=m\left(\frac{d^{2}B}{d\zeta^{2}}x_{0}^{2}+\frac{dB}{d\zeta}a-B_{0}\right),\label{eq:8}
\end{equation}
where we have assumed again that the magnet is anti-aligned with the
coil field. The upper case notation for the stiffness constant refers
to the complete float rather than an individual coil/magnet pair (as
compared to Equations~\ref{LinStiff} and~\ref{eq:6-1}). We can
see that in order to create a zero stiffness configuration we require
that the quantities in the bracket sum to zero. Noting that in our
configuration the second and third terms are both negative (see the
discussion above regarding the signs of the derivatives of the magnetic
field), in principle this is possible. However we did not pursue this
in the work described here but chose a value of $x_{0}$ and the positions
of the magnets relative to the coils in order to achieve a nominally
zero stiffness. The total rotational stiffness is given as the sum
of the terms in Equation~\ref{eq:8} from all the actuators, which
approximately multiplies it by a factor of eight. In Section~\ref{sec:Results}
we compare measurements of float's rotational stiffness with this
prediction.

We define the nominal centre-of-buoyancy (NCB) as
the point in the horizontal plane where the moments of the all the
forces acting on the float are zero when the coils carry their nominal
currents. If we consider the x direction, the position of the centre-of-buoyancy
with respect to the NCB of the float, $x_{cb}$,
can be modified by changing the stiffnesses of some actuators relative
to others. For example, using Equation~12 from Reference~13 we have
\begin{equation}
x_{cb}=\frac{K_{y\psi}}{K_{yy}}.\label{eq:9}
\end{equation}
The term in the denominator is the sum of the linear stiffnesses which
are given for each coil by Equation~\ref{LinStiff}. The term in
the numerator is a cross-term from the stiffness matrix describing
the static behaviour of the float and in the symmetrical case is zero.
We can compute the individual contributions from each coil/magnet
pair to this cross-term with the help of Equations~\ref{eq:6},~\ref{eq:7}
and~\ref{eq:8-1} to find 
\begin{equation}
k_{y\psi}=x_{0}m\frac{d^{2}B}{dy^{2}}.\label{CrossTerm}
\end{equation}
The change in centre-of-buoyancy in the $x$ direction is then, 
\begin{equation}
x_{cb}=\frac{K_{y\psi}^{a}+K_{y\psi}^{b}+K_{y\psi}^{e}+K_{y\psi}^{f}}{K_{yy}^{a}+K_{yy}^{b}+K_{yy}^{e}+K_{yy}^{f}},\label{eq:10}
\end{equation}
or 
\begin{equation}
x_{cb}\approx x_{0}\frac{\left(I^{b}+I^{e}-I^{a}-I^{f}\right)}{4I},\label{CoBchange}
\end{equation}
where $I$ is the average bias current in the coils labelled $a,b,e$ and $f$ in Figure~\ref{Setup} and clearly $x_{cb}$ is proportional to the difference in the currents flowing in the respective coils. This implies the theoretical maximum accuracy to which the centre-of-buoyancy can be tuned in the case presented here depends on the precision with which the actuators' strength can be changed, which in turn depends on the current noise of the coil drivers.

When the bearing is tilted by an angle, $\xi$, from the horizontal,
the total torque acting on the float from the actuators can then be
described by the following, 
\begin{equation}
\Gamma=\kappa\frac{\Delta I}{4I}+Mg\xi\left(x_{0}\frac{\Delta I}{4I}+x_{cm}\right),\label{FullTorque}
\end{equation}
where the first term in the brackets corresponds to $x_{cb}$ as given in Equation~\ref{CoBchange}, $M$ is the mass of the float, $g$ is the acceleration
due to gravity and $x_{cm}$ is the position of the float's centre-of-mass with respect to the NCB. Here $\kappa$ represents the torque due to any asymmetry
of the magnetic actuators and their positions around the float. In
the ideal case the process we employ for changing the bias currents
to tune the centre-of-buoyancy should not apply a torque,
but in any real system such a term does exist. Any instability in
the currents applied to the float to achieve a centre-of-buoyancy tuning
will introduce noise into the actual measurement via the $\kappa$ parameter,
so clearly it is desirable to reduce its magnitude as much as possible.
A similar expression to Equation~\ref{FullTorque} describes the centre
of mass tuning in the y direction.

In order to measure the $\kappa$ parameter and to check how the current
tuning shifts the centre-of-buoyancy we need to eliminate $x_{cm}$
from equation \ref{FullTorque}. We do this by adjusting the physical
centre-of-mass using balance weights, as described below, until it
coincides with the NCB. Then, dividing by the current
ratio, we find 
\begin{equation}
\tau=\Gamma/\frac{\Delta I}{4I}=\kappa+Mg\xi x_{0}.\label{Measurement}
\end{equation}
This equation states that, if the torque on the float is measured over a range
of tilt angles for a given set of bias currents in the coils, the
torque due to the actuator asymmetry can be calculated at $\xi=0$. Furthermore, it predicts a linear response of $\tau$ to the tilt angle. This gradient allows the change in centre-of-buoyancy described in Equation~\ref{CoBchange} to be experimentally determined. This comparison was verified by experiment. 

\section{\label{sec:Exp}Experimental Setup}

Our experimental setup consisted of 8 coil-magnet pairs; two on each
side of the square shaped float as shown in Figure~\ref{Setup}.
The float and bearing were made of aluminium alloy. The coils themselves
were based on the OSEM coils that have been developed for LIGO~\cite{Coils}
and each consisted of 500 turns of copper wire and had a mean radius
of 18 mm. The magnets used were grade N38 neodymium iron boron magnets
with a magnetic dipole moment of 0.775 N$\cdot$m/T and were cylindrical with
a radius and depth of 5 mm. They were attached to the float using
contact adhesive. The bearing's flat top surface consisted of 0.5
mm diameter holes in a 10 mm grid under the entire bottom surface
of the float through which compressed air was pumped at a constant
pressure to provide a lift force to the float.

\begin{figure}
\includegraphics[width=1\columnwidth]{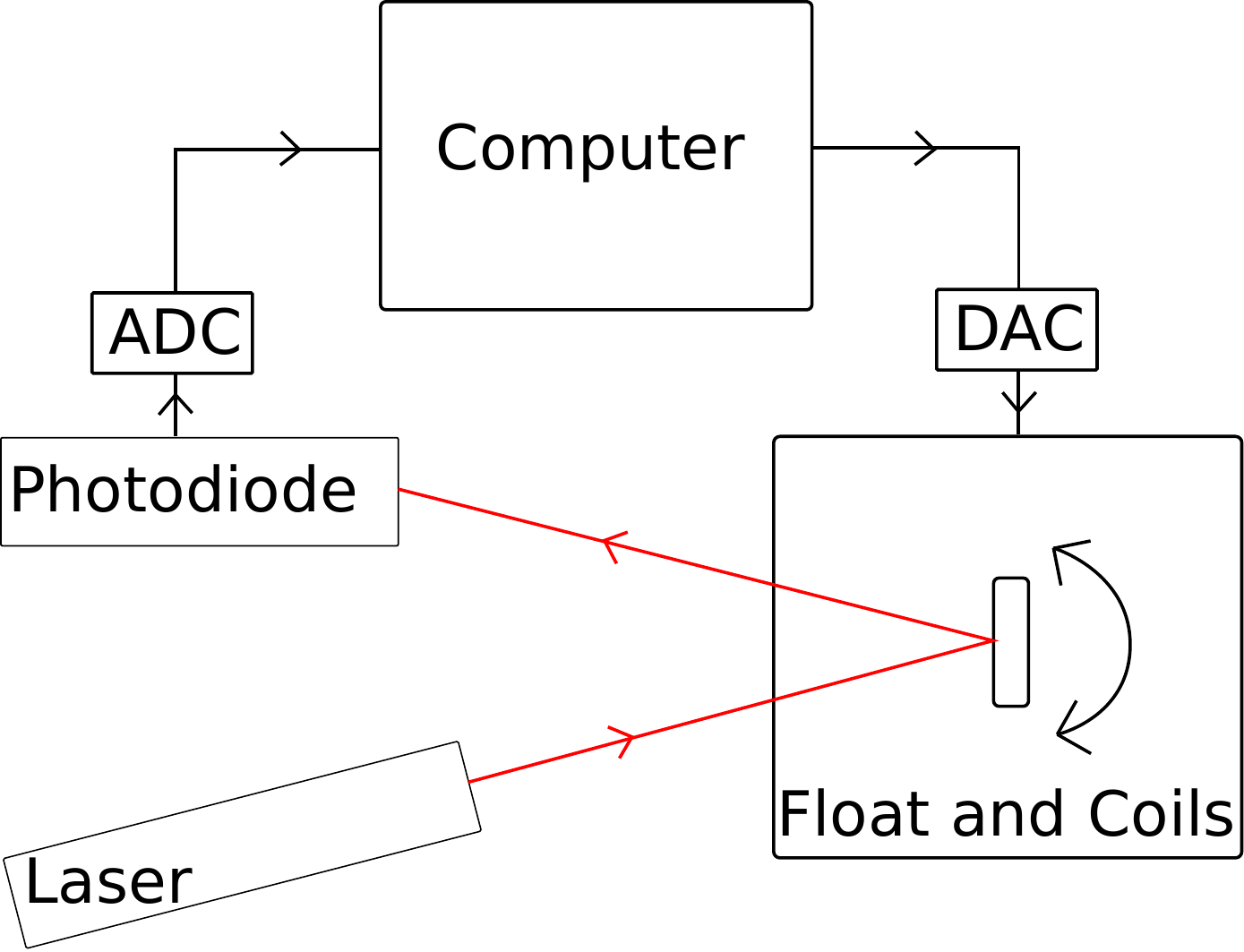}% Here is how to import EPS art
 \caption{\label{General} A general diagram of the full experimental setup.}
\end{figure}

Figure~\ref{Setup} gives a schematic drawing of the float, coils
and magnets, with a coordinate system centred on the float. A summary
of the dimensions of all the relevant components is given in Table~\ref{Dimensions},
\begin{table}[!htbp]
\centering \caption{A summary of the various dimensions of the float, bearing and magnetic
actuator setup. The dimension label corresponding to the equations
in Section~\ref{sec:Theory} is stated where applicable.}
\begin{ruledtabular}
\begin{tabular}{l r}
Parameter & Length (mm) $\pm$ 0.5 mm\\
\hline 
Coil-Magnet Axial Distance ($\zeta$) & $10.0$\\
Coil Mean Radius ($R$) & $18.0$\\
Coil-Magnet Radial Distance ($r$) & $0.0$\\
Coil Cross-Section Length & $8.0$\\
Coil Cross-Section Width & $9.0$\\
Actuator Arm Length ($x_{0}$) & $42.0$\\
Magnet Radius and Depth & $5.0$\\
Float Side Length & $115.0$\\
Float Depth & $10.0$\\
Bearing Tilt Length & $190.0$\\
Photodiode-Mirror Distance & $70.0$\\
\end{tabular}\label{Dimensions} 
\end{ruledtabular}
\end{table}
where the stated measurement uncertainty is used to propagate through
to the uncertainties on all measured torques and stiffnesses in Section~\ref{sec:Results}.
The rotation of the float was measured with an optical lever arrangement
with a laser reflecting off a small mirror attached to the centre
of the float and a position sensitive photodiode. This diode and its
associated electronic circuit was then connected to a computer via
an ADC. The computer was connected to the coils through a DAC. This
allowed the coil currents to be actively controlled via a PID control
loop in LabVIEW software and hence the float was kept stable relative
to a reference null position on the photodiode. A general diagram
of this setup is shown in Figure~\ref{General}.

\begin{figure}
\includegraphics[width=1\columnwidth]{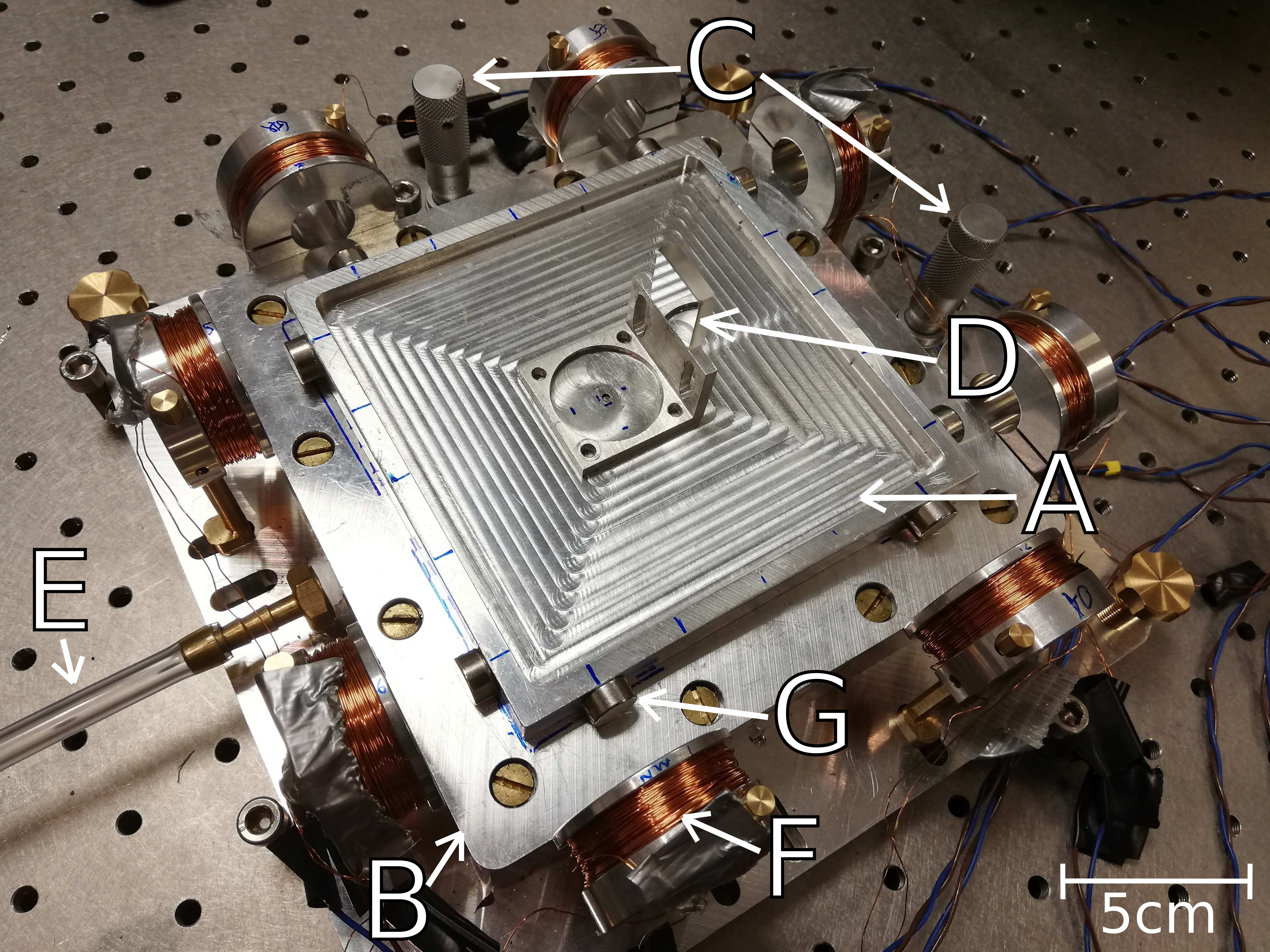}% Here is how to import EPS art
 \caption{\label{Real} A photograph of the setup; showing: A - float, B - bearing,
C - micrometers, D - mirror, E - compressed air input, F - a coil
and G - a magnet.}
\end{figure}

\begin{figure}
\includegraphics[width=1\columnwidth]{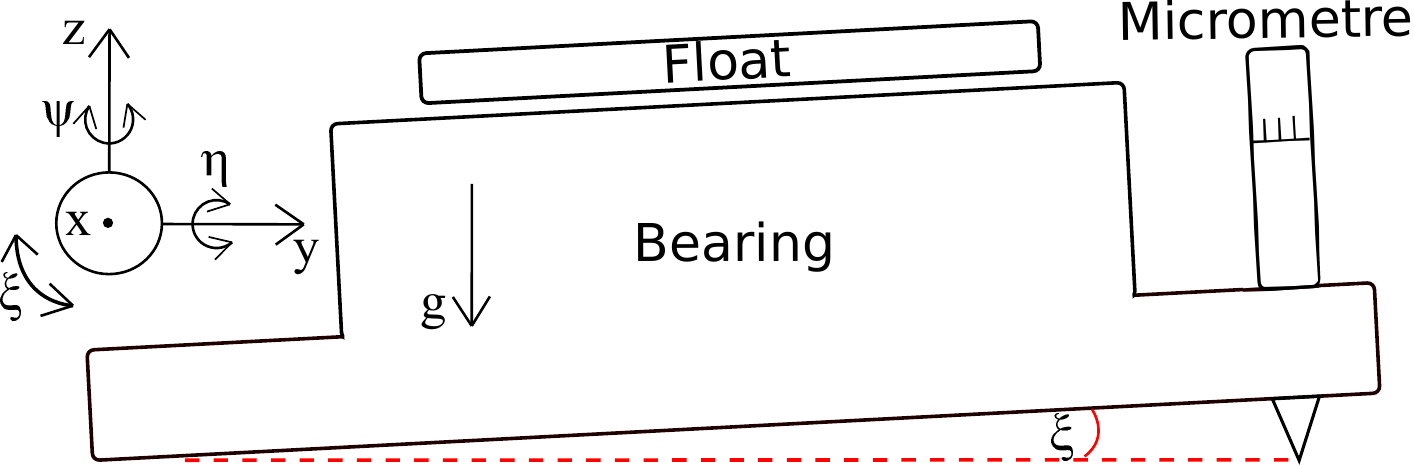}% Here is how to import EPS art
 \caption{\label{Tilt} A side-on schematic drawing showing the float, bearing
and the micrometre used to tilt the whole apparatus by an amount,
$\xi$, around the x-axis from the horizontal where $\xi=0$. The vertical is defined as the direction of the free-fall acceleration due to gravity, $g$. Positive and negative values of $\xi$ correspond to raising or lowering the setup from the horizontal respectively. The global coordinate system defined in Figure~\ref{Setup} is shown again here. The coils have been omitted
for clarity.}
\end{figure}

A photograph of the float, bearing, micrometers,
mirror, compressed air input, coils and magnets is shown in Figure~\ref{Real}.
Figure~\ref{Tilt} 
gives a schematic drawing of how the bearing can be tilted from the horizontal; where
this can be done along the float's x and y axes in the horizontal
plane. Tilt in the y-axis by a known tilt angle, $\xi$, is depicted
in the figure.

\section{\label{sec:Results}Results}

\begin{figure}
\includegraphics[width=1\columnwidth]{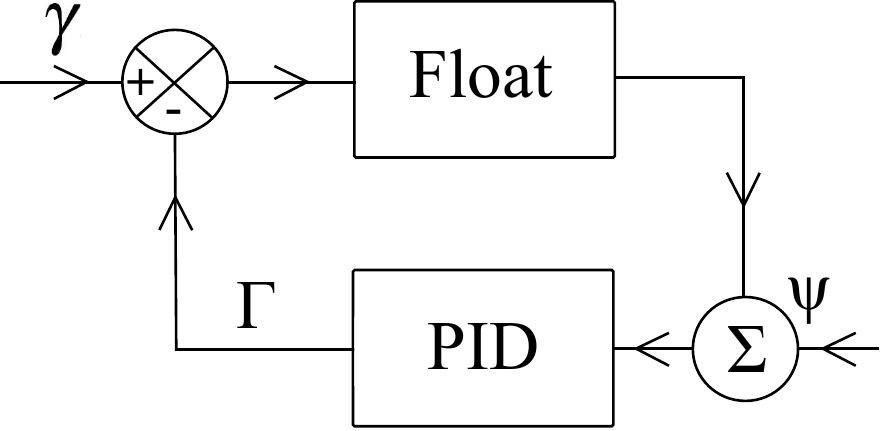}% Here is how to import EPS art
 \caption{\label{Servo} A block diagram showing the PID servo control. The
PID introduces a torque, $\Gamma$, to the float after a known offset
angle, $\psi$, is added to the controller input. There will also
be additional noise torques, $\gamma$, acting on the float.}
\end{figure}

The torque acting on the float could be measured
from the PID servo output, whose control loop is shown in Figure~\ref{Servo}.
The PID servo applied a torque, $\Gamma$, to the float by adding
or subtracting from the bias currents in the appropriate coils. All
eight coils were used for this purpose, and the bias currents are
those described in Equations~\ref{CoBchange},~\ref{FullTorque}
and~\ref{Measurement}. The magnitude of this change in current from
the bias currents could then be used in conjunction with Equation~\ref{Force}
multiplied by the actuator arm length term, $x_{0}$, from Table~\ref{Dimensions}
to give 
\begin{equation}
\Gamma=-8x_{0}m\frac{\partial B_{\zeta}}{\partial\zeta},\label{Torque}
\end{equation}
for eight coils. This measured torque is assumed to be equal to that
described by Equation~\ref{FullTorque}. The rotational stiffness
could be measured by recording the change in torque, $\Delta\Gamma$,
from Equation~\ref{Torque} applied by the PID servo to the float
after an offset equivalent to a known angle, $\Delta\psi$,
was added to the input of the controller. The rotational stiffness
would then be given by 
\begin{equation}
K_{\psi\psi}=-\frac{\Delta\Gamma}{\Delta\psi},\label{RotStiff}
\end{equation}
where this rotational stiffness is assumed to be equal to that described
by Equation~\ref{eq:8} once it had been summed over all the coils.
The linear transverse stiffness of the float in the x and y-axis of
the horizontal plane as described in Figure~\ref{Setup} could be
calculated using Equation~\ref{LinStiff} summed over the bias currents
of the four coils in each respective axis. As stated in Section~\ref{sec:Theory},
the positioning of the magnets was chosen to give $\theta=\pi$.

Before torque measurements could be made the float's centre-of-mass
displacement from its NCB, $x_{cm}$ from Equation~\ref{FullTorque},
had to be made zero. This was done by placing small masses on the
float in precise positions such that when tilting it from the horizontal
the PID servo torque did not change within the limit of the servo
readout noise. Doing this, while keeping the bias currents in all
the coils equal such that $\Delta I$ was zero in Equation~\ref{FullTorque},
implied that $x_{cm}$ was equal to zero. 

\begin{figure}
\includegraphics[width=1\columnwidth]{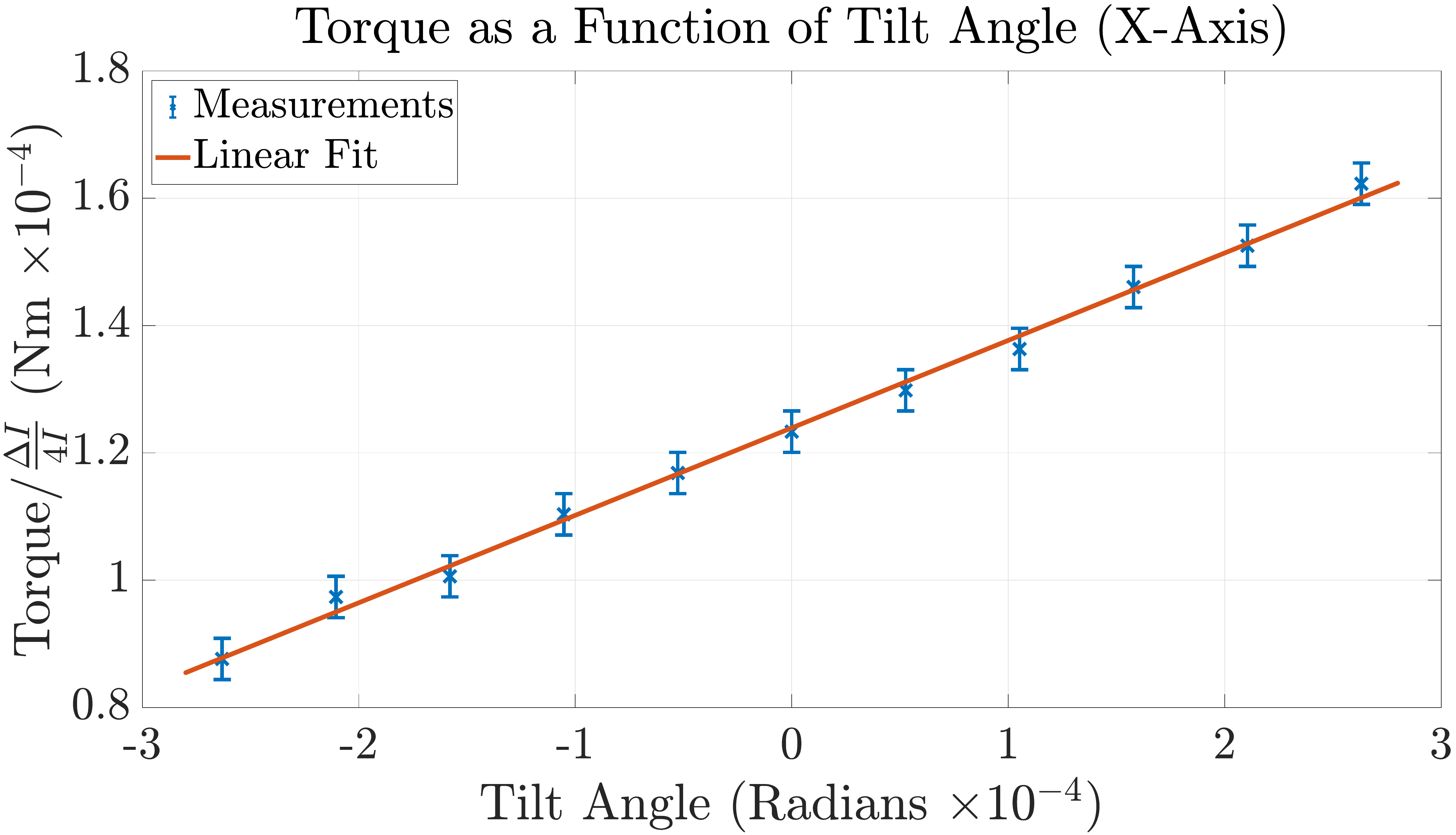}% Here is how to import EPS art
 \caption{\label{XDir} The torque, divided by the currents ratio term, acting
on the float over a range of tilt angles from the horizontal in the
x-axis.}
\end{figure}

\begin{figure}
\includegraphics[width=1\columnwidth]{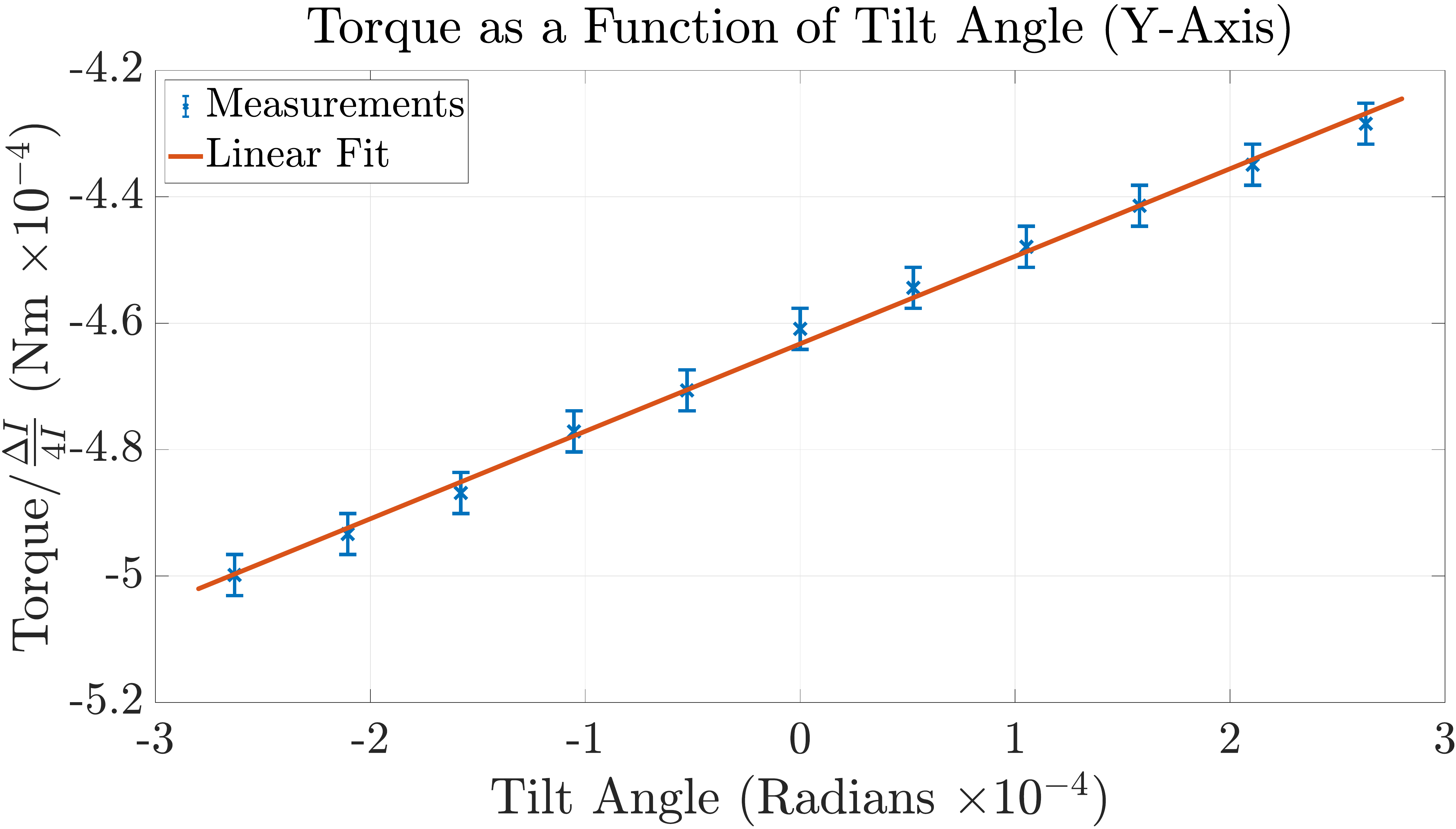}% Here is how to import EPS art
 \caption{\label{YDir} The torque, divided by the currents ratio term, acting
on the float over a range of tilt angles from the horizontal in the
y-axis.}
\end{figure}

The PID servo torque on the float was then measured over a range of
tilt angles from the horizontal, for a given set of bias currents.
This was done in both the x and y-axis of the float as shown in Figure~\ref{Setup}. The measurements in these axes are shown in Figures~\ref{XDir}
and~\ref{YDir}
respectively.

The total mass of the float, with the additional small masses used
for minimising $x_{cm}$, was 338.66 g. This, along with the value
of $x_{0}$ of (42.0 $\pm$ 0.5) mm from Table~\ref{Dimensions},
implied the expected gradient of the plots from Equation~\ref{Measurement}
should be (0.139 $\pm$ 0.003) N$\cdot$m/rad. The gradients from Figures~\ref{XDir}
and~\ref{YDir} are (0.137 $\pm$ 0.006) N$\cdot$m/rad and (0.138 $\pm$ 0.006)
N$\cdot$m/rad respectively. With the bias currents used and Equation~\ref{CoBchange},
the expected change of the centre-of-buoyancy of the float in both
the x and y-axis was (7.00 $\pm$ 0.17) mm. Using the gradients
from Figures~\ref{XDir} and~\ref{YDir}, in conjunction with Equations~\ref{CoBchange}
and~\ref{Measurement}, the measured changes in the centre-of-buoyancy
in the x and y-axis were calculated to be (6.90 $\pm$ 0.32) mm
and (6.95 $\pm$ 0.32) mm respectively. This shows
we had succeeding in tuning the float's centre-of-buoyancy to an accuracy
of $\pm$ 0.3 mm.

With the bias currents in all coils set to 0.075mA, and Equation~\ref{eq:8}
summed over all eight coils, the float's rotational stiffness was
calculated to be (-14.73 $\pm$ 0.41) $\mu$N$\cdot$m/rad. Using Equation~\ref{RotStiff} it was measured to be (-15.11 $\pm$ 2.05) $\mu$N$\cdot$m/rad. With these bias currents, the float's transverse stiffness in each horizontal axis, using Equation~\ref{LinStiff} summed over the four coils in
each axis, was calculated to be (1.11 $\pm$ 0.16) N/m. This gave
a natural oscillation frequency in each axis of (0.29 $\pm$ 0.04) Hz. All the results are summarised in Table~\ref{Summary}.
\begin{table}[!htbp]
\centering \caption{A summary of the measurement results.}
\begin{ruledtabular}
\begin{tabular}{l r}
Attribute  & Result\\
\hline 
\multicolumn{2}{c}{Plot Gradients}\\
\hline 
Float x-Axis & (0.137 $\pm$ 0.006) N$\cdot$m/rad\\
Float y-Axis & (0.138 $\pm$ 0.006) N$\cdot$m/rad\\
Equation~\ref{Measurement} Prediction & (0.139 $\pm$ 0.003) N$\cdot$m/rad\\
\hline 
\multicolumn{2}{c}{Centre-of-Buoyancy Change}\\
\hline 
Float x-Axis & (6.90 $\pm$ 0.32) mm\\
Float y-Axis & (6.95 $\pm$ 0.32) mm\\
Equation~\ref{CoBchange} Prediction & (7.00 $\pm$ 0.17) mm\\
\hline 
\multicolumn{2}{c}{Float Rotational Stiffness}\\
\hline 
Measurement & (-15.11 $\pm$ 2.05) $\mu$N$\cdot$m/rad\\
Equation~\ref{eq:8} Prediction & (-14.73 $\pm$ 0.41) $\mu$N$\cdot$m/rad\\
\hline 
\multicolumn{2}{c}{Float Transverse Stiffness and Frequencies}\\
\hline 
Equation~\ref{LinStiff} x and y-Axis Calculation & (1.11 $\pm$ 0.16) N/m\\
Oscillation Frequency x and y-Axis & (0.29 $\pm$ 0.04) Hz\\
\end{tabular}\label{Summary} 
\end{ruledtabular}
\end{table}

\section{\label{sec:Disc}Discussion}

The measurements in Figures~\ref{XDir} and~\ref{YDir} change linearly
with tilt angle as expected from Equation~\ref{Measurement}. The
gradients of these plots were expected to be 0.139 N$\cdot$m/rad and the measured
gradients all lie within $1.5\%$ of this value and within their uncertainty
ranges. The calculated and measured changes in the centre-of-buoyancy
displacement of the float from Figures~\ref{XDir} and~\ref{YDir}
were also within $1.5\%$ of each other and within each others uncertainty
ranges. Thus all the results are in excellent agreement with the theory.
Crucially this demonstrates that it is possible to tune in-situ the
centre-of-buoyancy of a suspended object and hence also possible to
decouple its rotational motion from ground tilt and horizontal accelerations.

The measured and calculated rotational stiffnesses of the float were
within $2.5\%$ of each other and within their respective uncertainty
bounds. The magnitude of these values is an order of magnitude lower
than the rotational stiffness of a torsion balance used in a recent
determination of Newton's constant of gravitation, which had a stiffness
of approximately 218 $\mu$N$\cdot$m/rad~\cite{BigG}. Additional actuators could
increase the float's transverse stiffness from the calculated values
to make the float more transversely stable, and also allow the adjustment
of its rotational stiffness. Preliminary measurements with this setup
show that this is possible~\cite{Zero}. The natural oscillation
frequencies of the float in the transverse plane were calculated.
With the float being levitated through an air bearing, it can be assumed
that it's motion was highly damped. As such in this case there was
no risk of the oscillations significantly affecting the float's motion.

The design could be improved to allow greater precision in the placement and adjustment of the different components to give more accurate results. The noise from the electronics of the positional photodiode was measured to be $8.4\times10^{-7}$ N$\cdot$m/$\sqrt{\mathrm{Hz}}$, while the total noise of the air bearing system was measured to be $2.7\times10^{-6}$ N$\cdot$m/$\sqrt{\mathrm{Hz}}$. This was the largest contribution to the data point uncertainties in Figures~\ref{XDir} and~\ref{YDir}. This gave an error on the gradients of these plots, which was then propagated onto the experimental estimation of $x_{cb}$ from Equation~\ref{CoBchange}. No attempt was made to put the device in a protective enclosure or shield the photodiode from environmental light sources. As such the error on $x_{cb}$ of $\pm$ 0.3 mm could be reduced by lowering the noise of the air bearing. This could be done by using a different method of levitation other than a pressurised-air suspension, such as an electrostatic or superconducting suspension~\cite{Zero}. From Equation~\ref{CoBchange}, the theoretical maximum accuracy of centre-of-buoyancy tuning that could be attained with the setup presented here, taking into account the current noise from the coil drivers of the magnetic actuators averaged over a second, is  $\pm$ $1.6\times10^{-8}$ m. This corresponds to a possible improvement in accuracy of a factor of over 18000. 

\section{\label{sec:Conc}Conclusion}

An air bearing suspension that levitates a float, with its motion
in the horizontal plane of the laboratory controlled using magnetic
actuators, was constructed. The observed behaviour of the float was
compared to the predictions of a detailed model of the statics of
the float-actuator system and they were found to be consistent. The
results from Figures~\ref{XDir} and~\ref{YDir} demonstrate the
in-situ electromagnetic tuning of the float's centre-of-buoyancy to
an accuracy of $\pm$ 0.3 mm. This result implies it is practical
to decouple the rotational mode of a suspended object from tilt and
horizontal accelerations due to seismic noise, by tuning its centre-of-buoyancy
to lie at its nominal centre-of-buoyancy (NCB).

This result paves the way for other, more sensitive, experiments to
be designed with a view to performing weak-force measurements at sub-mm
ranges. Work is ongoing on a superconducting torsion balance~\cite{Zero,Float}.
The aim here is to develop an instrument which exhibits the same advantages
as the air bearing where it can be tuned in-situ to be rotationally
decoupled from ground tilt, in addition to allowing the in-situ tuning
of its rotational stiffness. This combined with a transverse stiffness
provided by the superconducting magnetic actuators should allow measurements
of the inverse square law of gravity down to mass separations of the
order of 10$\mu$m. The Newtonian torque signal,
given a day's integration, requires a fundamental noise level of less
than $1\times10^{-14}$ N$\cdot$m/$\sqrt{\mathrm{Hz}}$. Given a typical seismic noise
acceleration spectral density of $5\times10^{-7}$ m/s$^{2}\sqrt{\mathrm{Hz}}$
and a suspended mass of 338.66 g, we would need to match the centre-of-buoyancy
and mass to an accuracy of about $1\times10^{-7}$ m. So an improvement
in tuning accuracy of a factor of around a thousand would be required
compared with what is achieved here. The uncertainty on the matching
of the centres-of-mass and buoyancy is limited by the overall
noise in the air bearing system, so this goal may be achievable with
a superconducting or other type of suspension. We should also mention
that a possible downside of this technique is the way that the actuation
system can introduce noise into the measurement through its asymmetries
(the $\kappa$ parameter introduced in Equation~\ref{FullTorque}).
Any strategy of reducing this factor would include making such asymmetries
as small as possible in the first instance, thus ensuring that the
actuation torques themselves are as small as possible.

\begin{acknowledgements} We would like to thank Dr. Chris Collins
for helpful discussions. We would also like to thank Masters project students Joseph Parker, Jack Joynson, Hugo Yamaguchi and Arthur O'Leary for their help in this
project. We would additionally like to thank John Bryant and David Hoyland in helping to design and build the PID servo system in LabVIEW and the ADC/DAC setup for the experiment. We are grateful to UK STFC (Grant No. ST/F00673X/1) who initially supported this work. We are also very grateful to Leverhulme (Grant No. RPG-2012-674) for financial support.
%We wish to acknowledge the support of the author community in using
%REV\TeX{}, offering suggestions and encouragement, testing new versions,
%\dots.
\end{acknowledgements}

\nocite{*}
\bibliography{Draft}
 % Produces the bibliography via BibTeX.
\end{document}